\documentclass[prb, twocolumn, showpacs]{revtex4}

\usepackage{bm}
\usepackage{epsfig, amsmath,amssymb}

\begin{document}

\title{Geometry Induced Potential on a 2D-section of a Wormhole: Catenoid}

\author{Rossen Dandoloff}
\affiliation{ Laboratoire de Physique Th\'{e}orique et
Mod\'{e}lisation, Universit\'{e} de Cergy-Pontoise,
 F-95302 Cergy-Pontoise, France}
 \email{rossen.dandoloff@u-cergy.fr}

\author{Avadh Saxena}
\affiliation{Theoretical Division and Center for Nonlinear Studies,
Los Alamos National Laboratory, Los Alamos, NM 87545 USA}
\email{avadh@lanl.gov}

\author{Bj\o rn Jensen}
\affiliation{Vestfold University College, \\ Faculty of Science and Engineering, N-3103 
T\o nsberg, Norway} 
\email{bjorn.jensen@hive.no} 

\begin{abstract}
We show that a two dimensional wormhole geometry is equivalent to a
catenoid, a minimal surface.  We then obtain the curvature induced
geometric potential and show that the ground state with zero energy 
corresponds to a reflectionless potential.  By introducing an appropriate 
coordinate system we also obtain bound states for different angular 
momentum channels.  Our findings can be realized in suitably bent bilayer 
graphene sheets with a neck or in a honeycomb lattice with an array of 
dislocations or in nanoscale waveguides in the shape of a catenoid. 

\end{abstract}


\maketitle

\section{Introduction}
Quantum mechanics in flat two dimensional space gives unusual results such 
as the quantum anticentrifugal force for waves with zero angular momentum 
\cite{qm1, qm2}.  Thus, it would be insightful to explore quantum mechanics in 
curved two dimensional space.   Of special interest are the minimal surfaces 
(i.e. with zero mean curvature) which play an important role in physics.  Besides 
the plane, the two other examples include the helicoid and the catenoid.  
An interesting question in the context of the three dimensional wormhole 
geometry \cite{worm1, worm2} in cosmology is whether information can 
propagate across the wormhole. We study the analog of this  problem in two 
dimensions and first show that the two-dimensional wormhole is a catenoid.  
We then obtain the curvature induced quantum potential \cite{dacosta}.   The 
latter is an attractive geometric potential $V_G(q_1,q_2)=-(\hbar^2/8m_0)(\kappa_1 
-\kappa_2)^2$, where $\kappa_1$, $\kappa_2$ denote the two position-dependent 
principal curvatures of the surface, $(q_1,q_2)$ are the surface coordinates, 
and $m_0$ is the mass of the particle on the surface.   

A two-dimensional wormhole geometry can conceivably be realized in a bilayer 
of honeycomb lattices with radially arranged dislocations or in bilayer graphene  
\cite{joglekar}, where the curvature induced suppression of local Fermi energy 
can lead to the control of local electronic properties.  In the next section we 
demonstrate the equivalence of a two-dimenasional wormhole and a catenoid. 
In Sec. III we obtain the effective curvature induced potential.  In Sec.  IV we 
introduce a suitable coordinate system to study the bound states of the resulting 
Schr\"odinger equation on the catenoid.   We have previously studied bound 
states on the other minimal surface, namely the helicoid \cite{atanasov}.  Note 
that a genus one helicoid also exists \cite{hoffman}.  Finally, in Sec. V we 
summarize our main conclusions and comment on the anticentrifugal force.

\section{Catenoid as a two-dimensional section of a wormhole}
For a catenoid $x=R\cosh(z/R)\cos\phi$, $y=R\cosh(z/R)\sin\phi$ and
$z=z$ with $\phi \in [0,2\pi]$ (Fig. 1). Thus the local radius $\rho=R\cosh(z/R)$ 
and the metric is given by the following elements:
\begin{equation}
g_{\rho\rho}=\frac{\rho^2}{\rho^2-R^2} , ~~~~g_{\phi\phi}=\rho^2 .
\end{equation}

\begin{figure}[ht]
\begin{center}
   \includegraphics[scale=0.4]{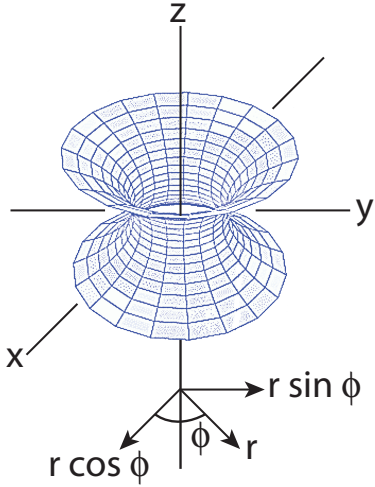}
   \caption{\label{neck} {A two-dimensional section (catenoid) of a three 
   dimensional worm hole geometry with its axis along $z$ and the throat 
   radius $R$.
}}
\end{center}
\end{figure}

We now show that a two dimensional wormhole geometry is equivalent to
a catenoid (Fig. 1).  In cylindrical coordinates $(z,r,\phi)$ a
two-dimensional section of a wormhole is given by
\begin{equation}
z(r)=\pm b_0 \ln\left[\frac{r}{b_0}+\sqrt{\frac{r^2}{b_0^2}-1}\right] ,
\end{equation}
with $l=\pm\sqrt{r^2-b_0^2}$.  For the three dimensional wormhole the line element is
given by the following expression\cite{worm2}:
\begin{equation}
ds^2 = dl^2 + (b_0^2+l^2)(d\theta^2+\sin^2\theta d\phi^2), 
\end{equation}
where the coordinates belong to the following intervals: $l\in[-\infty,+\infty]$, $\theta\in[0,\pi]$ and $\phi\in[0,2\pi]$ and $b_0$ is the shape function of the wormhole [in general $b=b(l)$ and for $l=0, b=b(0)=b_0=const$ represents the radius of the throat of the wormhole].  Here $l$ is a radial coordinate measuring proper radial distance; $\theta$ and $\phi$ are spherical polar coordinates.
In this paper we will consider the case $\theta=\pi/2$ which represents an equatorial section
of a three dimensional wormhole (at constant time). For this section we thus get the following
line element:
\begin{equation}
ds^2 = dl^2 + (b_0^2+l^2)d\phi^2 ,
\end{equation}
which is precisely equivalent to the line element of a catenoid (since $l^2=r^2-b_0^2$)
\begin{equation}
ds^2=\frac{r^2}{r^2-b_0^2}dr^2 + r^2 d\phi^2 .
\end{equation}

Note that if we consider any other section of the three dimensional wormhole, say for $\theta=\theta_0$
the line element will change to:
\begin{equation}
ds^2=\frac{r^2}{r^2-b_0^2}dr^2 + a^2r^2 d\phi^2 .
\end{equation}
where $a^2=\sin^2\theta_0$ and obviously $a^2\in[0,1]$. For the catenoid this will mean only a rescaling of the radius of the catenoid from $R$ to $aR$. The line element Eq. (6) corresponds
to a catenoid with $x=aR\cosh(z/aR)\cos\phi$, $y=aR\cosh(z/aR)\sin\phi$ and
$z=z$. Thus all $\theta$-sections of the 3D-wormhole represent a catenoid with radius $aR$. The catenoid with the biggest radius corresponds to the equatorial section $\theta=\frac{\pi}{2}$ and 
with zero radius to $\theta=\pi$.

\section{Effective Potential} 

Returning to the catenoid and focusing on the $(z,\phi)$ coordinates
(instead of $\rho$, $\phi$), the line element is given by
\begin{equation}
ds^2=\cosh^2(z/R)dz^2+R^2\cosh^2(z/R)d\phi^2 ,
\end{equation}
with the principal curvatures
\begin{equation}
\kappa_1=\frac{1}{R}{\rm sech}^2(z/R), ~~~~\kappa_2=-\frac{1}{R}{\rm
sech}^2(z/R) .
\end{equation}
This implies that the mean curvature $H=(\kappa_1+\kappa_2)/2=0$ (i.e. a minimal 
surface) and the Gaussian curvature $K=\kappa_1\kappa_2=-(1/R^2){\rm sech}^4(z/R)$.
If a particle is confined to move on a curved surface (with finite thickness) and the thickness 
is allowed to go to zero, then there will appear an effective potential in the Schr\"odinger 
equation, known as the da Costa potential \cite{dacosta}.  (For a flat surface the potential 
is zero).  The corresponding curvature induced da Costa potential for a catenoid is
\begin{equation}
V(z)=-\frac{\hbar^2}{2m_0}(H^2-K)=-\frac{\hbar^2}{2m_0R^2}
{\rm sech}^4(z/R) .
\end{equation}

Note that for $a^2\ll 1$ the geometric potential becomes very deep and localized at the origin.

The relevant Schr\"odinger equation is
\begin{eqnarray}
&-&\frac{\hbar^2}{2m_0R\cosh^2(z/R)}\left[R\frac{\partial^2\psi}{\partial
z^2}+\frac{1}{R}\frac{\partial^2\psi}{\partial\phi^2}\right] \nonumber\\
&-&\frac{\hbar^2}{2m_0R^2}{\rm sech}^4(z/R)\psi=E\psi ,
\end{eqnarray}
or simplifying

\begin{eqnarray}
-R\frac{\partial^2\psi}{\partial z^2}-\frac{1}{R}\frac{\partial^2\psi}
{\partial\phi^2}-\frac{{\rm sech}^2(z/R)}{R}\psi=\nonumber\\
\frac{2m_0R}{\hbar^2}\cosh^2(z/R)E\psi .
\end{eqnarray}

Using the cylindrical symmetry along the $z$-axis, we set $\psi=e^{im z}\Phi$
and we get the following equation for $\Phi$:
\begin{equation}
\Phi_{zz}-\frac{m^2}{R^2}\Phi+\frac{{\rm sech}^2(z/R)}{R^2}\Phi
+\frac{2m_0E\cosh^2(z/R)}{\hbar^2}\Phi=0 .
\end{equation}

Defining dimensionless length $\zeta=z/R$ and energy $\epsilon=2m_0E
R^2/\hbar^2$ we get the following effective Schr\"odinger equation:
\begin{equation}
-\Phi_{\zeta\zeta}+V(\zeta)\Phi(\zeta) = 0 ,
\end{equation}
where the geometric potential now reads:
\begin{equation}
V(\zeta)=[m^2-\epsilon\cosh^2(\zeta)] - {\rm sech}^2(\zeta) .
\end{equation}
This potential for $m\neq 0$ bears some similarity to the corresponding
geometric potential for the three dimensional wormhole \cite{worm1}. 
Note that in the ground state ($\epsilon=0$, called also a {\it critically bound state} 
\cite{Lekner}) the above potential becomes the reflectionless Bargmann's potential 
\cite{bargmann} and the Schr\"odinger equation becomes the hypergeometric 
equation with the ground state wavefunction (or the Goldstone mode) given by 
$\Phi(\zeta)={\rm sech}(\zeta)$. This result is remarkable in that the minimal surface 
of catenoid enables complete transmission across it for a quantum particle. This 
does not seem to be the case for a three dimensional wormhole.  For nonzero and
positive $\epsilon$ the above potential is an inverted double well
potential shown in Fig. 2.

\begin{figure}[ht]
\begin{center}
   \includegraphics[scale=0.4]{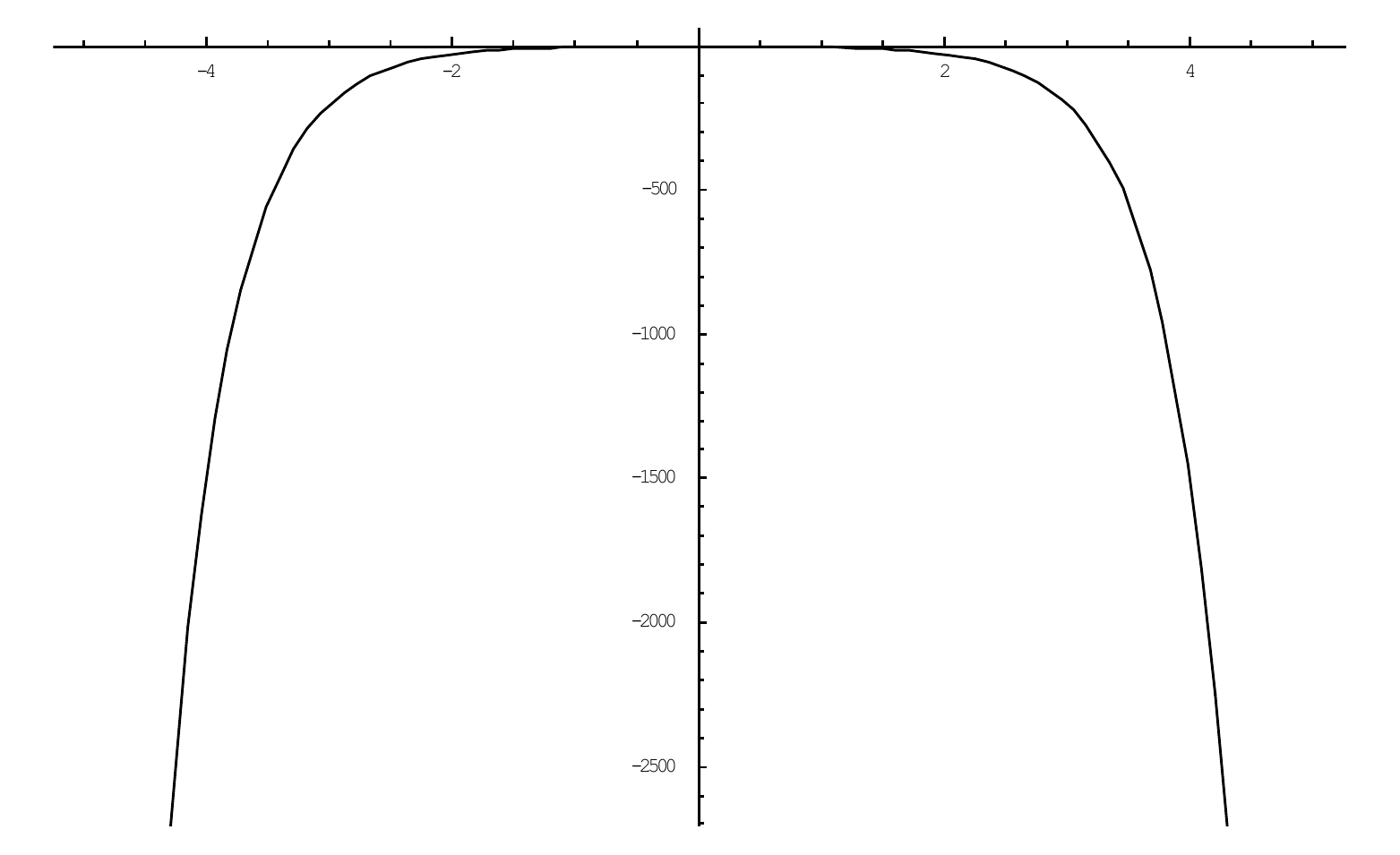}
   \caption{\label{neck} {The inverted double well potential $V(\zeta)$ with $m=\pm 1$ and
    $\epsilon =0.1$.
}}
\end{center}
\end{figure}

\section{Bound States} 

Let us consider Eq. (13) in more detail. We see that
\begin{equation}
\lim_{\zeta\rightarrow\pm\infty}|V|\rightarrow\infty\, .
\end{equation}
The behavior of the potential at infinity is strange since the physical geometry in the catenoid in these regions approaches the usual Euclidean one. This feature can be traced to the coordinates used since the proper length per unit in the $\zeta$ direction diverges when $\zeta\rightarrow\pm\infty$. This can be remedied by introducing another set of coordinates on the catenoid.

Quantum theory in curved spaces is generally a challenge since the theory is not generally covariant.  Classical quantum theory is not even Lorentz invariant.  This puts a severe constraint on the coordinate system in which one wishes to describe the physics in order to be able to extract the physical content of the theory.  This challenge was even central in the early days of general relativity theory itself in connection with the physical interpretation of the Schwartzshild metric, e.g. just like as in general relativity theory one is usually safe concerning the physical interpretation as well as the definitions of physical quantities when the manifold in question is asymptotically Minkowski (Euclidean). In such asymptotic regions we expect on physical grounds to rederive the usual flat space physical results. The asymptotic properties thus in some sense anchor the curved region and its physics to reality as we know it.  The catenoid is an asymptotic Euclidean object, 
thus making this manifold a space anchored to ``reality".

Considering the 2D Schr\"odinger equation in the plane in polar coordinates we get the Bessel equation.  Clearly, the boundary condition at the origin is suspect here.  However, in our case we can as a first approximation consider a deformation of the plane in a region around the origin. In the deformed region the Schr\"odinger equation will generally be very complicated but the solutions of it must nevertheless be matched to the Bessel functions which survive sufficiently far from the deformed region.  This reasoning goes {\it ad verbum} through also on the catenoid even though we here, in addition to curvature corrections, also have a topology change when compared to the plane.  Hence, we should seek coordinates on the catenoid such that the Schr\"odinger equation gives rise to the Bessel equation in the asymptotic region on the catenoid.  The coordinates should in particular result in a metric which is reminiscent of polar coordinates at infinity. It is possible to find such coordinates if one covers the entire manifold with two coordinate patches.  One patch covers the region $\zeta>0$ and the other one $\zeta<0$.  In the upper part we choose
\begin{equation}
\eta^+=e^\zeta-1\, ; \, \zeta>0\, .
\end{equation}
In the lower part we correspondingly choose
\begin{equation}
\eta^-=-(e^{-\zeta}-1)\, ;\, \zeta<0\, .
\end{equation}
Clearly $\eta^+=\eta^-$ at $\zeta=0$.  The invariant line-element can then be written as
\begin{equation}
ds^2=\frac{((\eta^\pm\pm 1)^2+1)^2}{4(\eta^\pm\pm 1)^4}(d\eta^\pm )^2+\frac{1}{4}(\frac{(\eta^\pm\pm 1)^2+1}{\eta^\pm\pm 1})^2d\phi^2\, ,
\end{equation}
In the limit $\eta^\pm\rightarrow \pm\infty$ the metric reduces to
\begin{eqnarray}
ds^2=\frac{1}{4}(d\eta^\pm )^2+\frac{1}{2}((\eta^\pm\pm 1)^2+1)d\phi^2 \nonumber\\
\simeq\frac{1}{4}(d\eta^\pm )^2+\frac{1}{2}(\eta^\pm )^2 d\phi^2 \, .
\end{eqnarray}
Hence, the asymptotic form of this metric is very similar to the usual polar coordinates.  Clearly, these new coordinates should be well suited to explore the physical states of a quantum particle on the catenoid.

Let us now consider the Schr\"odinger equation. In terms of the new coordinates we have in particular that
\begin{eqnarray}
&&\partial^2_u\Phi = (\eta^\pm\pm 1)\partial_\pm((\eta^\pm\pm 1)\partial_\pm\Phi)\, , \\
&&\cosh u=\pm\frac{1}{2}(\frac{(\eta^\pm\pm 1)^2+1}{\eta^\pm\pm 1})\, .
\end{eqnarray}
This gives rise to identical expressions for the Schr\"odinger equation in the two patches. In the upper patch the equation is explicitly given by
\begin{eqnarray}
\partial^2_+\Phi +\frac{1}{(\eta^++1)}\partial_+\Phi +\bigg[\frac{\epsilon}{4}-\frac{(m^2-\epsilon /2)}{(\eta^++1)^2}+ \nonumber\\
\frac{1}{4}(\frac{\epsilon}{(\eta^++1)^4}+\frac{16}{((\eta^++1)^2+1)^2})\bigg]\Phi =0\, .
\end{eqnarray}
Clearly, letting $\eta^+\rightarrow\infty$ we get the Bessel equation, which is well behaved at infinity.

The stationary Schr\"odinger equation, and assuming a well defined energy $E$ eigenvalue problem, is formally given by
\begin{equation}
(-\nabla^2+V)\Psi =E\Psi\, .
\end{equation}
Hence, we have that
\begin{eqnarray}
V-E= -\bigg[\frac{\epsilon}{4}-\frac{(m^2-\epsilon /2)}{(\eta^++1)^2}+ \nonumber\\
\frac{1}{4}\left(\frac{\epsilon}{(\eta^++1)^4}+\frac{16}{((\eta^++1)^2+1)^2}\right)\bigg]\, .
\end{eqnarray}
In the asymptotic region we find that
\begin{equation}
\lim_{\eta^+\rightarrow\infty}V=E-\frac{1}{4}\epsilon >0\, .
\end{equation}
We have plotted the potential for $m=0,\pm 1,\pm 2$ and $\epsilon =2$ in  Fig. 3.

\begin{figure}[ht]
\begin{center}
   \includegraphics[scale=0.5]{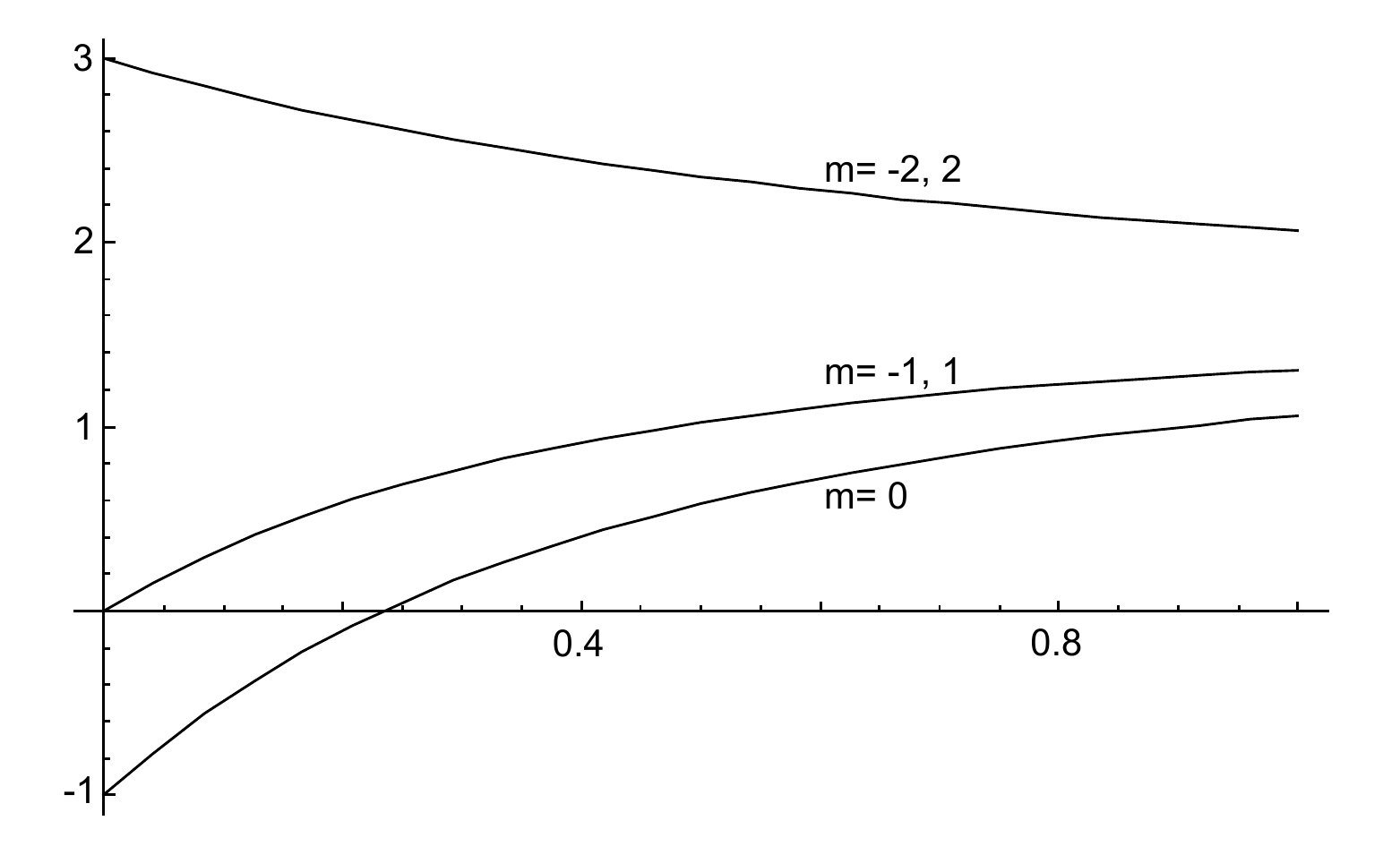}
   \caption{\label{neck} {The  potential $V(\eta^+)$ with $m=0,\pm 1, \pm 2$ and $\epsilon =2$.
}}
\end{center}
\end{figure}

Clearly, the constant part of the potential can be renormalized to zero without any physical consequences. Hence, the renormalized potential $V_r$ can be taken to be
\begin{eqnarray}
V_r-E= -\bigg[-\frac{(m^2-\epsilon /2)}{(\eta^++1)^2}+ \nonumber\\
\frac{1}{4}\left(\frac{\epsilon}{(\eta^++1)^4}+\frac{16}{((\eta^++1)^2+1)^2}\right)\bigg]\, .
\end{eqnarray}
Consider $V$ and the case when the energy is set to unity.  Then the physical quantum states fall into four different categories.  In the $s$-channel ($m=0$) the potential becomes negative sufficiently close to the origin.  When $m=\pm 1$ the same patterns emerge but with a much faster fall off of the potential with increasing coordinate distance from the origin than in the $s$-channel. When $m=\pm 2$ the potential is everywhere positive definite.  Higher angular momentum modes will give rise to bound states at a distance from the origin and at a distance in the direction of increasing radial coordinate.

In general, if it is possible to redefine $\cosh^2(\zeta)\Phi_{\zeta\zeta}=\Phi_{\eta\eta}$
in Eqs. (13) and (14) then the equation for $\Phi(\eta)$ would correspond to the double 
sinh-Gordon potential which is quasi-exactly solvable \cite{razavy,dshg}, i.e. for specific 
values of $\epsilon$ and $m$ exact solutions can be obtained.

\section{Conclusion}
We demonstrated that a two dimensional equatorial ($\theta=\frac{\pi}{2})$ or any other  
section of a wormhole is equivalent to the minimal surface of a catenoid. We then showed
 that the curvature induced da Costa quantum potential \cite{dacosta} allows for a critically bound state $(\epsilon=0)$. This leads to a reflectionless transmission of a quantum particle across the 
 catenoid (or a 2D wormhole).  By introducing an appropriate coordinate system we were able to obtain bound states for different angular momentum channels.  It is interesting to note that the potential is attractive at the origin $\eta^{\pm}=0$ for $m=0$  and $m=\pm 1$ (anticentrifugal potential with bound states). In contrast, in the plane the anticentrifugal potential is present only  
for $m=0$ but an additional $\delta$-function potential is needed at the origin in order to 
introduce the missing length scale in the plane\cite{qm1}. We note that a radial array of 
dislocations in a bilayer of honeycomb lattices or a suitably bent bilayer graphene sheet 
with a neck \cite{joglekar} may provide a physical realization for our findings.   Another experimental means of measuring the potential (Eq. 14) is to construct nanoscale waveguides 
in a  catenoid shape. 

This work was supported in part by the U.S. Department of Energy.

\end{document}